\documentclass[useAMS, usenatbib, letter]{aa}  
\usepackage{graphicx}
\usepackage{txfonts}
\usepackage{graphicx,lscape}
\usepackage{amssymb,amsmath}
\usepackage{rotating}
\usepackage{xcolor}
\usepackage{caption}
\usepackage{subcaption}

\def\msun{M_\odot}
\def\fs{f_{\mathrm S}}

\def\I0st{{I_{\mathrm 0}^{\mathrm{st}}}}
\def\V0{V_{\mathrm 0}}

\def\t0{t_{\mathrm 0}}

\def\u0{u_{\mathrm 0}}

\def\piE{\pi_{\mathrm{E}}}

\def\piEE{\pi_{\mathrm{EE}}}
\def\piEN{\pi_{\mathrm{EN}}}

\def\muS{\mu_\mathrm{S}}

\def\thetaE{\theta_{\mathrm{E}}}

\begin{document} 

\title{Is there a nearby microlensing stellar remnant hiding in $Gaia$ DR3 astrometry?}
\titlerunning{Is there a microlensing remnant in Gaia DR3 astrometry?}
\authorrunning{M. Jab{\l}o{\'n}ska et al.}

   \author{
   Maja Jab{\l}o{\'n}ska\inst{1}
   \and
   {\L}ukasz Wyrzykowski\inst{1}
   \and
   Krzysztof A.~Rybicki\inst{1,2}
   \and
   Katarzyna Kruszy{\'n}ska\inst{1}
   \and
   Zofia Kaczmarek\inst{3}
   \and
   Zephyr Penoyre\inst{3}
}          

\institute{
Astronomical Observatory, University of Warsaw, Al.~Ujazdowskie~4, 00-478~Warszawa, Poland,
\and
Department of Particle Physics and Astrophysics, Weizmann Institute of Science, Rehovot 76100, Israel,
\and
Institute of Astronomy, University of Cambridge, Madingley Road, CB3 0HA, Cambridge, UK
}

   \date{July 2022}

  \abstract  
  {  Massive galactic lenses with large Einstein Radii should cause a measurable astrometric microlensing effect, i.e. the light centroid shift due to the motion of the two images. Such a  shift in the position of a background star due to microlensing was not included in the $Gaia$ astrometric model, therefore significant deviation should cause $Gaia$'s astrometric parameters to be determined incorrectly.
Here we studied one of the photometric microlensing events reported in the $Gaia$ DR3, GaiaDR3-ULENS-001, for which poor goodness of $Gaia$ fit and erroneous parallax could indicate the presence of the astrometric microlensing signal. Based on the photometric microlensing model, we simulated $Gaia$ astrometric time-series with the astrometric microlensing effect added. We found that including microlensing with the angular Einstein Radius of $\theta_{\rm E}$ = $2.60^{+0.21}_{-0.24}$ mas ($2.47^{+0.28}_{-0.24}$ mas) assuming positive (negative) impact parameter $u_0$ reproduces well the astrometric quantities reported by $Gaia$.
We estimate the mass of the lens to $1.00^{+0.23}_{-0.18}$ $\msun$ ($0.70^{+0.17}_{-0.13}$ $\msun$) and its distance to $0.90^{+0.14}_{-0.11}$ kpc ($0.69^{+0.13}_{-0.09}$ kpc), proposing the lens could be a nearby isolated white dwarf.
}
 

   \maketitle
%

\section{Introduction}



$Gaia$ Data Release 3 (GDR3) contained 1.806 billion sources down to G$\sim$21 mag, with 1.46 billion sources with full astrometric solution \citep{EDR3}. The model fit to those sources had five or six parameters: position (two parameters), proper motions in RA and Dec (two parameters), parallax, and, in fainter cases, the sixth parameter - pseudocolour - is added \citep{Lindegren2021}. Such a model describes an isolated source with linear motion on the sky perturbed by the $Gaia$'s orbital motion around the Sun. Any additional motion (e.g. due to unresolved binarity) was not included in the standard model, therefore sources with more complex motion will have the model not matching well \citep{Lindegren2012, Belokurov2020, penoyre2020binary, Lindegren2021}. 
GDR3 catalogue contains a number of statistics on the goodness of the astrometric fit to the individual sources, e.g. parallax\_over\_error, $\chi^2$, astrometric excess noise (AEN), or astrometric excess noise significance. \cite{Lindegren18RUWE} introduced RUWE (Renormalized Unit Weight Error), being the empirically re-normalized Unit Weight Error (UWE), which is the square root of reduced $\chi^2$. The re-normalization takes into account the brightness and colour of the source. \cite{Lindegren18RUWE} strongly advised to use RUWE instead of UWE or other astrometric model quality measures derived in Gaia data releases in order to identify sources with astrometry not matching the 5 or 6-parameter model. For example, in the works of \cite{penoyre2020binary} and \cite{Belokurov2020} it is shown that RUWE statistic can be used to identify unresolved binary stars in the GDR2 data set as the photo-centre wobble causes additional residuals to the basic astrometric 5-parameter model. There were recently also other applications of astrometric statistics to identify astrometric centroid deviations, e.g. \cite{Bouma2020}, \cite{Gomel2020} and \cite{Gandhi2020}.


Astrometric microlensing can be another cause of an additional apparent motion in the sources hiding in the GDR3 data. This effect (\citealt{Hog1995}, 
\citealt{Miyamoto1995}, \citealt{Walker1995}, 
\citealt{Dominik2000}) is caused by the appearance of two lensed images of the source (in the case of a single lens) and their evolution in position and brightness while the source and the lens change their relative position projected on the sky. Since the separation between the images in the case of Galactic microlensing is of the order of milliarcseconds, the images are typically not resolvable. However, the centre of light shifts during the event and the amplitude of the shift is comparable to the separation between the images. $Gaia$ has been long predicted to be able to detect such anomalies \citep{BelokurovEvans2002, Rybicki2018, Kluter2020} as the mission is expected to have enough accuracy to detect subtle changes of position at the sub-mas level. Detection of astrometric microlensing leads to a direct measurement of the angular Einstein radius ($\theta_\mathrm{E}$), which is necessary to measure the mass and distance of the lens from the equations:
\begin{equation}
    M_{\rm L}=\frac{\theta_\mathrm{E}}{\kappa \pi_\mathrm{E}}, \quad
D_{\rm{L}}=\frac{1}{\theta_\mathrm{E} \pi_{\rm{E}} + 1/D_{\rm{S}}}\, ,
\label{eq:mass}
\end{equation}
where $\kappa=4G/(c^2~\mathrm{au}) = 8.144~\mathrm{mas}/\msun$,  $D_\mathrm{S}$ is the distance to the source in kpc
and $\piE$ is the unit-less microlensing parallax, caused by the non-linear motion of the observer along the Earth's orbital plane around the Sun. The effect of microlensing parallax often causes subtle deviations and asymmetries in the light curves of events lasting a few months or more, so that the Earth's orbital motion cannot be neglected \citep{Gould2000b}.


$Gaia$ is expected to publish all its individual astrometric, photometric and spectroscopic data points for all sources in its fourth data release (GDR4)\footnote{https://www.cosmos.esa.int/web/gaia/release}. In GDR3, the only astrometric information for sources with no non-single-star solution is embedded in the 5/6-parameter model of linear proper motion and parallax.
In this work we searched for potential signatures of astrometric microlensing in the GDR3 astrometric statistics for microlensing events reported by $Gaia$ itself in the first catalogue of microlensing events \citep{Wyrzykowski2022DR3}. We identify the best candidate event, which is bright enough to have the astrometric microlensing signal detectable and is located away from the Bulge, where other factors could affect $Gaia$'s astrometry.  We investigate the possibility that astrometric deviations reported by $Gaia$ are caused by astrometric microlensing.  Basing on simulations of $Gaia$ astrometric data in combination with $Gaia$'s photometry, we estimate the angular size of the Einstein Radius $\theta_{\rm E}$ to be in the range inferred from the two solutions: $2.60^{+0.21}_{-0.24}$ mas (positive $u_0$) and $2.47^{+0.28}_{-0.24}$ mas (negative $u_0$) and conclude that the lens could be a $1.00^{+0.23}_{-0.18}$ $\msun$ ($0.70^{+0.17}_{-0.13}$ $\msun$) isolated object, most likely a Main Sequence or a white dwarf star, at a distance of $0.90^{+0.14}_{-0.11}$ kpc ($0.69^{+0.13}_{-0.09}$ kpc) from the Sun.

\section{Selection}

\cite{Wyrzykowski2022DR3} identified 363 microlensing events in $Gaia$ DR3 data covering the years 2014-2017 located all over the sky. 
The vast majority, naturally, was located within the Galactic Plane, in particular towards the Galactic Bulge.
From that sample, we selected events with baseline magnitudes brighter than G$<$16 mag in $Gaia$ $G$-band. This selection was motivated by the study of \cite{Rybicki2018}, who simulated $Gaia$ astrometric microlensing events and estimated the brightness limit at which $Gaia$’s astrometric accuracy is sufficient to detect astrometric microlensing signal. 

Here, in the selection process, we removed events located towards and close to the Galactic Bulge, within 20 deg in Galactic longitude. Despite having the highest sky-density of events, the $Gaia$ astrometric data in that part of the sky is likely to suffer from multiple errors, often unaccounted in the $Gaia$ data processing, caused by crowding and source confusion. 


We used RUWE parameter as an indicator of potential microlensing signal in the $Gaia$ astrometric data and limited our sample to sources with RUWE$>$2 \citep{Lindegren18RUWE}. 
In our selection process, we also decided to choose bright microlensing events from DR3 with G$_0$$<$16 mag and there was only one event, satisfying this and the aforementioned criteria, GaiaDR3-ULENS-001= 6059400613544951552.

\section{Microlensing model and mass estimate in GaiaDR3-ULENS-001 event}

GaiaDR3-ULENS-001 (RA=184.4362°, Dec=-59.0294°), is a microlensing event which reached its maximum brightness at G=12.783 mag in 2015, but went unnoticed by then on-going microlensing surveys \citep{Wyrzykowski2022DR3}.
It is located at Galactic coordinates (298.6005°, 3.5587°)
which is in the Crux constellation in the Southern hemisphere.

\begin{figure}
    \centering
    \includegraphics[width=9cm]{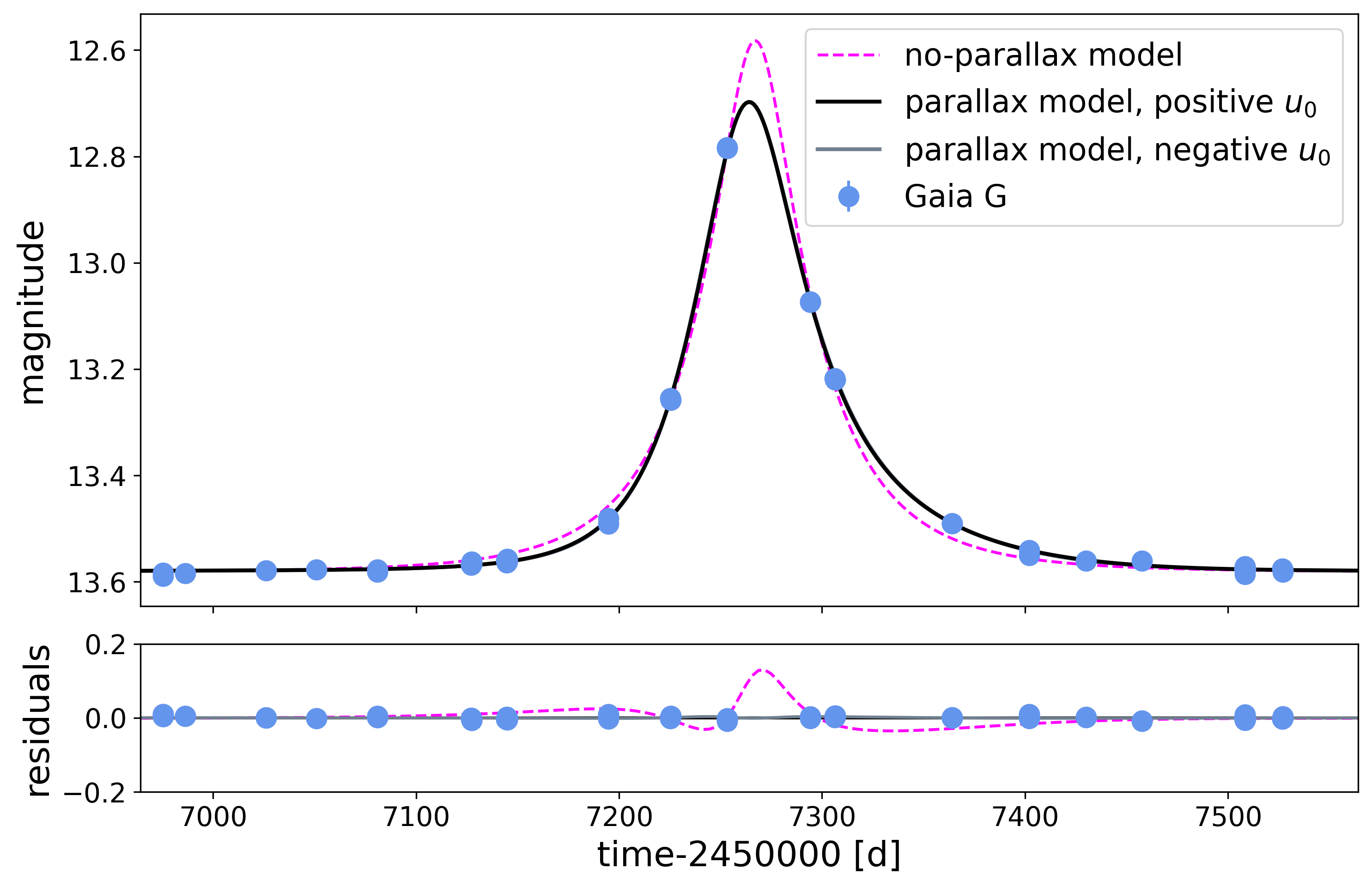}
    \caption{GaiaDR3-ULENS-001 event light curve with the best parallax models for both solutions (solid dark lines) and standard model without the parallax (dashed magenta line). Both parallax solutions overlap and are indistinguishable on this plot.
    }
    \label{fig:lc}
\end{figure}

Photometric data of GaiaDR3-ULENS-001 in $G$, $G_{BP}$ and $G_{RP}$ $Gaia$ bands were obtained from the $Gaia$ DR3 data archive.
The data were fit with standard and parallax microlensing models using \texttt{MulensModel} open source software \citep{mulensmodel}. The parallax model included space-parallax due to the fact that $Gaia$ spacecraft is located in the L2 point of the Sun-Earth system, at 1.5 million km from Earth. However, the space-parallax signal was practically not detectable in the $Gaia$ data.
Figure \ref{fig:lc} shows the photometric data with the best models found. 
The MCMC algorithm implemented in the \texttt{emcee} package \citep{foreman2013emcee} identified two solutiond in the annual parallax model due to classical degeneracy in the impact parameter and parallax vector (the lens can pass the source on two sides). Their parameters are listed in Table \ref{tab:photo_model}. Both solutions indicate there is little light coming from objects other than the source.

\begin{table}[]
    \centering
\def\arraystretch{1.3}
\begin{tabular}{r|c|c}
parameter & positive $u_0$ solution & negative $u_0$ solution \\
\hline
$t_0$ [JD] & 2457264.11$^{+0.27}_{-0.28}$ & 2457264.17$^{+0.30}_{-0.31}$ \\
$t^{helio}_{\rm E}$ [d] & 47.14$^{+4.18}_{-3.85}$ & 45.65$^{+5.65}_{-4.77}$ \\
$u_0$ & 0.51$^{+0.08}_{-0.07}$ & -0.51$^{+0.07}_{-0.09}$ \\
$\pi_{\rm E,N}$ & 0.31$^{+0.05}_{-0.05}$ & 0.42$^{+0.07}_{-0.06}$ \\
$\pi_{\rm E,E}$ & -0.08$^{+0.07}_{-0.07}$ & -0.03$^{+0.10}_{-0.10}$ \\
$G_0$ [mag] & 13.58$^{+0.01}_{-0.01}$ & 13.58$^{+0.01}_{-0.01}$ \\
$f_S$ & 1.11$^{+0.29}_{-0.22}$ & 1.10$^{+0.30}_{-0.23}$ \\
\hline
$\chi^2/dof$ & 58.29/60 & 65.66/60 \\
\end{tabular}
    \caption{Two solutions found for the photometric model of GaiaDR3-ULENS-001 with microlensing parallax ($\piEN, \piEE$) and blending ($\fs$, source to total flux ratio in $Gaia$ G band). $\t0$ is the moment of the minimal lens-source distance at $\u0$ (in Einstein Radii units). $G_0$ is the baseline brightness in the $Gaia$ G band.}
    \label{tab:photo_model}
\end{table}

The values obtained for the photometric model of GaiaDR3-ULENS-001 were used to generate the simulated astrometric microlensing signal. We used the \texttt{astromet}\footnote{https://github.com/zpenoyre/astromet.py} Python package to generate the astrometric track, which was then also used to simulate the $Gaia$ 5-parameter fit \citep{penoyre2022astrometric}. Information about the observation epochs is provided by the \texttt{scanninglaw} package \citep{scanninglaw}. 

The photometric model provides all parameters for the astrometric model except the angular Einstein Radius $\thetaE$, the source distance $\varpi$ and source proper motion $\muS$. The direction of the relative proper motion between the lens and the source ($\mu_{\rm rel}=\thetaE/t_{\rm E}^{\rm helio}$) is also known from photometry as it is parallel with the vector of the parallax $\vec{\pi_{\rm E}}$ \citep{Gould2000b}.

\begin{figure}
    \centering
    \begin{subfigure}[b]{.45\textwidth}
         \centering
         \includegraphics[width=9cm]{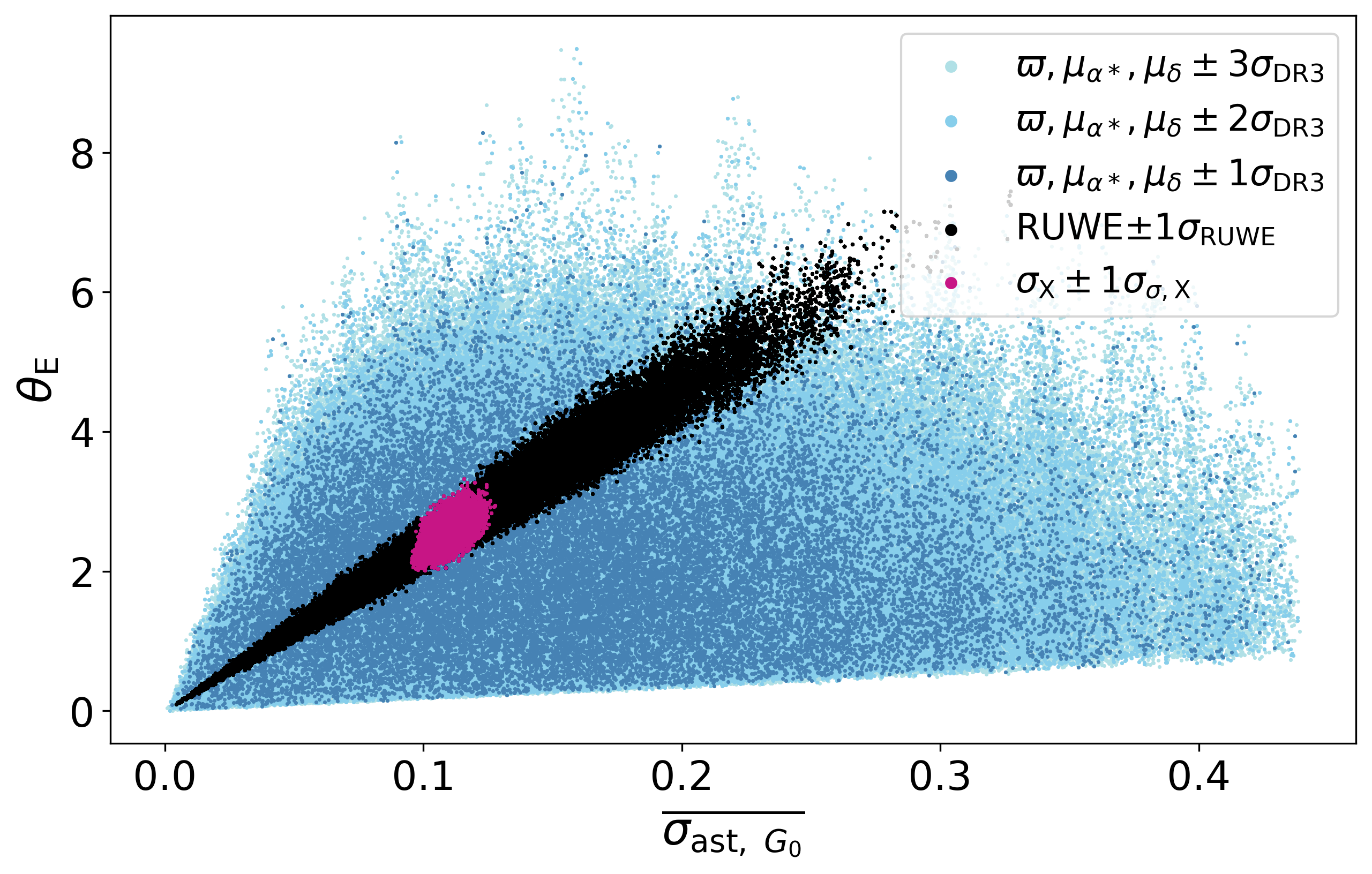}
         \caption{positive $u_0$ solution}
    \end{subfigure}
    \vfill
    \begin{subfigure}[b]{.45\textwidth}
         \centering
         \includegraphics[width=9cm]{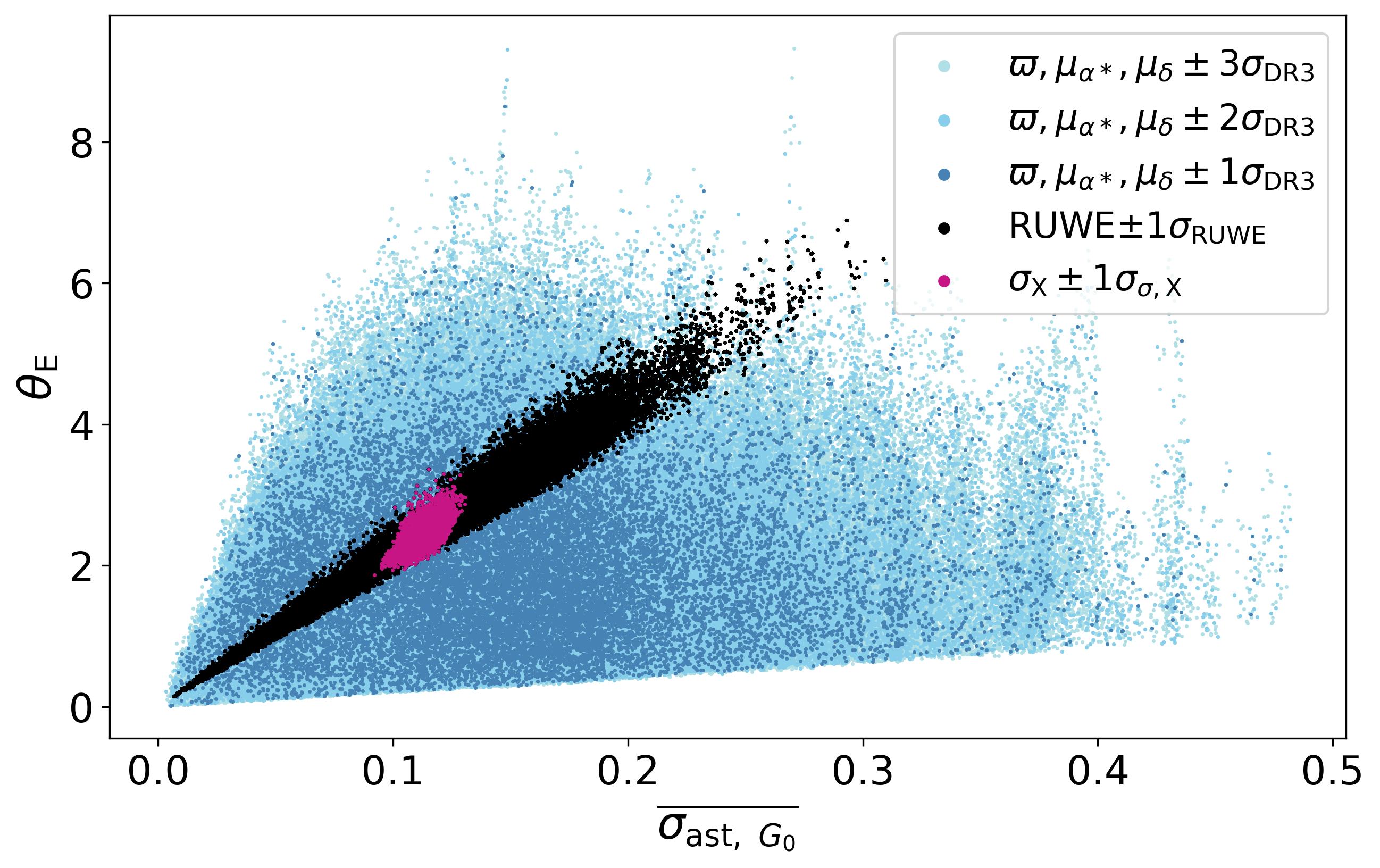}
         \caption{negative $u_0$ solution}
    \end{subfigure}
    \caption{Einstein Radius ($\thetaE$) used in the astrometric microlensing data simulation for GaiaDR3-ULENS-001 which reproduces $Gaia$ DR3 5-parameter astrometric fit to within their 3$\sigma$ (grey circles), as a function of mean astrometric per-CCD error-bar ($\overline{\sigma_{ast,\ G_0}}$) for the base magnitude $G_0$. Black circles mark those values, which also reproduce value of RUWE to within $1\sigma_{RUWE}\approx$ 0.12. Pink circles mark values which additionally reproduce parallax and proper motion errors to within $1\sigma_{\sigma,X}\approx$ 0.0018. X denotes the parameters of $\varpi$, $\mu_{\alpha*}$, and $\mu_\delta$, as their error values exhibit an almost linear correlation and very close values for every sample.}
    \label{fig:theta-sigma}
\end{figure}

The input parameters for the astrometric model are ($\varpi$, $\mu_{\alpha*}$, $\mu_{\delta}$, $\theta_\mathrm{E}$). 
The AL errors for the specific astrometric measurements used in GDR3 are unknown, as the only available errors measured are the astrometric $\chi^2$ and RUWE. Therefore, in order to simulate the errors to fit them to the 5-parameter solution we estimated them by using the mean error function of the measured magnitude adopted from \cite{Lindegren2021}. However, the errors can still differ for every measurement, as the function provides just the mean. Instead of applying a random function to the errors, we have decided on introducing the scaling parameter $\sigma_\mathrm{AL}$ by which the sum of simulated errors is multiplied. This approach results in the same behaviour for time-series with different number of observations and modifies the sum of AL errors, which is actually included in the final \textit{Gaia}'s 5-parameter fit. To constrain the error scale, we filter the results not only by RUWE, but also by $\sigma_\varpi$, $\sigma_{\mu,\alpha*}$, and $\sigma_{\mu,\delta}$, since they are also dependent on the astrometric errors. The dependence of the $\theta_E$ parameter satisfying the DR3 values on the mean astrometric per-CCD error-bar that would be obtained for the baseline magnitude in the $G$ band assuming some value of $\sigma_\mathrm{AL}$ from the function presented by \cite{Lindegren2021} can be seen in Figure \ref{fig:theta-sigma}. Since the parameters $\sigma_\varpi$, $\sigma_{\mu,\alpha*}$, and $\sigma_{\mu,\delta}$ have very similar values for a given set of astrometric model input parameters and an almost linear correlation, on the plot they are treated as one parameter and denoted as $\sigma_{\sigma, X}$.

\textit{Gaia}'s data contain 527 AL observations with 523 observations marked as good. Our simulator returns 585 AL observations. This is not alarming, as not all of the observations predicted based on the scanning law are published due to instrumental reasons, such as an unsatisfactory quality of the measurements or lack of space for all the measurements in dense fields \citep{scanninglaw}.

To fit the parameters to reproduce the actual Gaia DR3 5-parameter solution within its $3\sigma$ we have used the MCMC algorithm implemented in \texttt{NumPyro} package \citep{numpyro_bingham2019pyro, numpyro_phan2019composable}. To include the photometric model uncertainties, each sample was generated assuming random values from a normal distribution centered on the values presented in Table \ref{tab:models}. We have then constrained the results by RUWE and values of $\sigma_\varpi$, $\sigma_{\mu,\alpha*}$, $\sigma_{\mu,\delta}$. The photometric scatter caused the astrometric model to be non-deterministic in regards to the input parameters. The accepted range of RUWE and $\sigma_X$ values were estimated by bootstrapping the astrometric model samples and resulted in $\sigma_{RUWE}\approx$ 0.12 and $\sigma_{\sigma,X}\approx$ 0.0018.

We found that astrometric data with $\thetaE$ in range from 2.23 mas to 2.81 mas was fit with DR3 values within $3\sigma$, RUWE within $1\sigma_{RUWE}$ and parallax and proper motion uncertainties within $1\sigma_{\sigma,X}$.

With the model including microlensing signal we also derived the source parameters, which differ from the ones in the 5-parameter astrometric model without microlensing. The resulting values for the positive $u_0$ solution are: $\varpi=0.28^{+0.05}_{-0.05}$ mas, $\mu_{\alpha*}=-6.42^{+0.08}_{-0.06}$ mas/yr, and $\mu_{\delta}=1.44^{+0.08}_{-0.08}$ mas/yr, while for the negative $u_0$ solution: $\varpi=0.37^{+0.07}_{-0.06}$ mas, $\mu_{\alpha*}=-6.22^{+0.10}_{-0.08}$ mas/yr, and $\mu_{\delta}=1.48^{+0.06}_{-0.07}$ mas/yr.

\begin{figure}
    \centering
    \begin{subfigure}[b]{.5\textwidth}
         \centering
         \includegraphics[width=9cm]{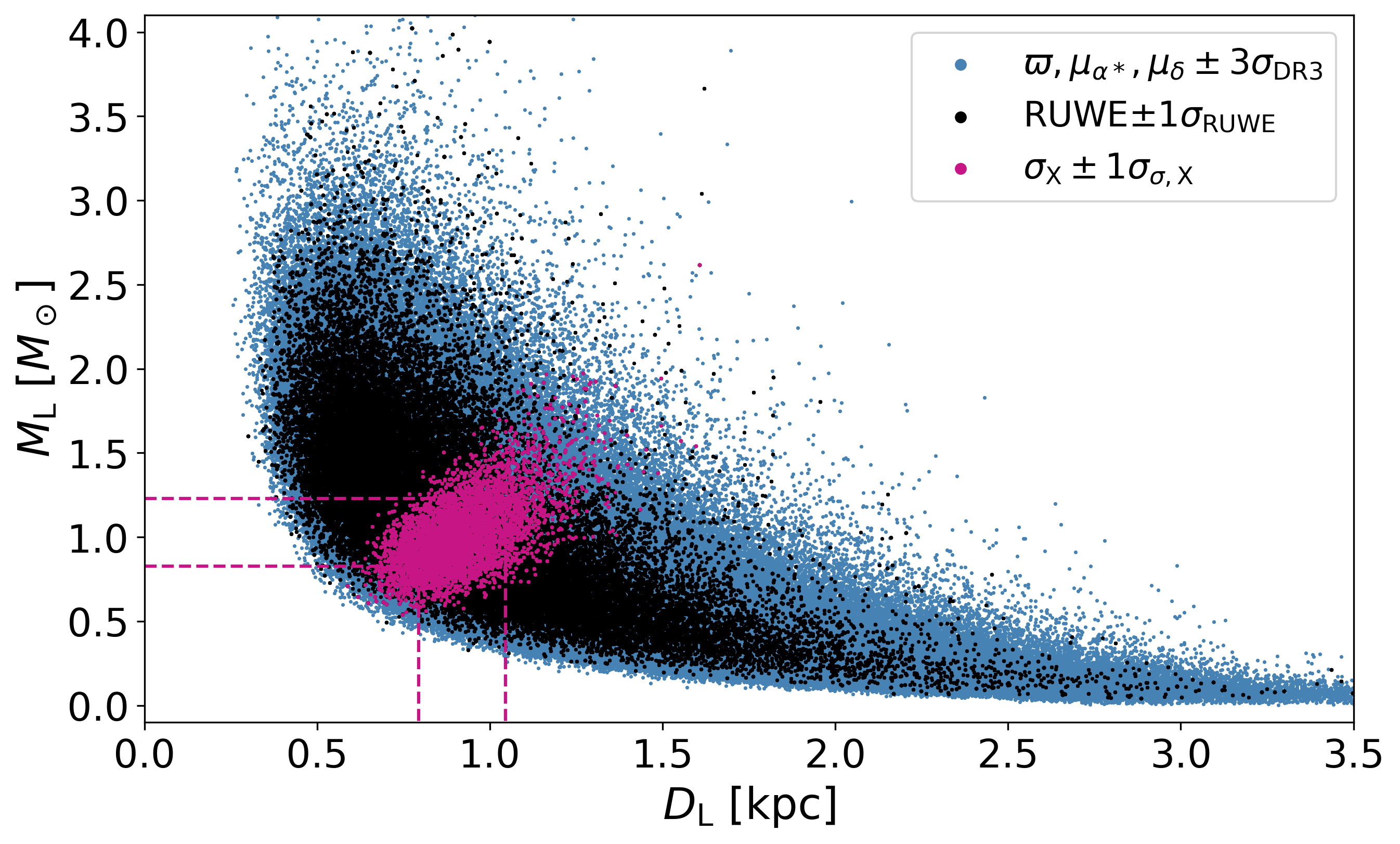}
    \end{subfigure}
    \vfill
    \begin{subfigure}[b]{.5\textwidth}
         \centering
         \includegraphics[width=9cm]{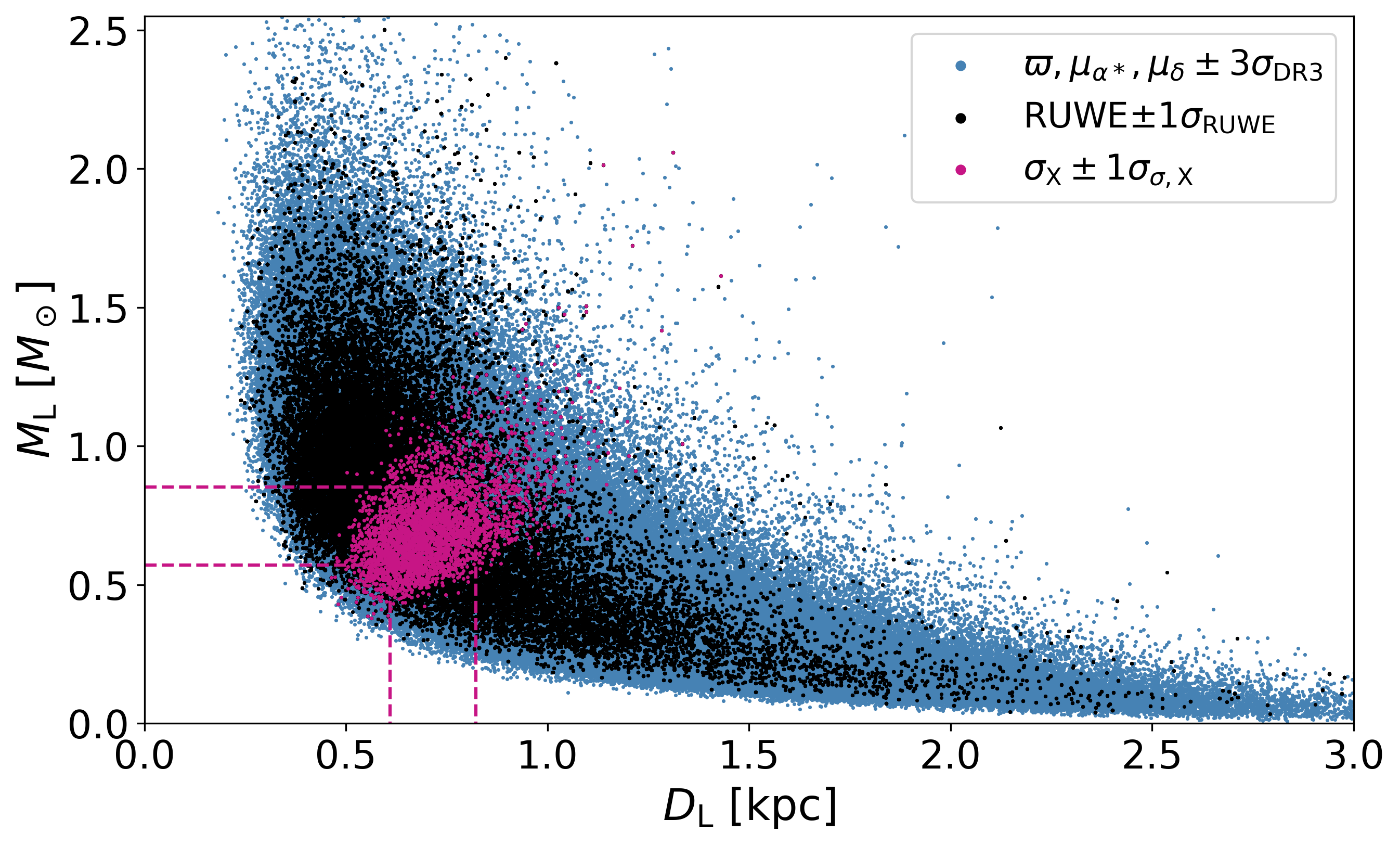}
    \end{subfigure}
    \caption{Inferred lens mass versus lens distance for GaiaDR3-ULENS-001 for the positive $u_0$ (top panel) and negative $u_0$ (bottom panel) models. The lines indicate the $1\sigma$ bounds.}
    \label{fig:ml-dl}
\end{figure}

Figure \ref{fig:ml-dl} shows the mass and distance values inferred from the simulated samples using Equations \ref{eq:mass} for both microlensing solutions. 
 Combining the microlensing parallax value obtained from the DR3 photometry and the  $\theta_\mathrm{E}$ values which reproduce $Gaia$ DR3 data, we estimated the mass of the lens as in the ranges of $1.00^{+0.23}_{-0.18}$ $\msun$ ($0.70^{+0.17}_{-0.13}$ $\msun$) and its distance in the ranges of $0.90^{+0.14}_{-0.11}$ kpc ($0.69^{+0.13}_{-0.09}$ kpc).

\begin{table}[]
    \centering
\def\arraystretch{1.3}
\begin{tabular}{r|c|c}
parameter & positive $u_0$ solution & negative $u_0$ solution \\
\hline
$M_L$ [$M_\odot$] & $1.00^{+0.23}_{-0.18}$ & $0.70^{+0.17}_{-0.13}$ \\
$D_L$ [kpc] & $0.90^{+0.14}_{-0.11}$ & $0.69^{+0.13}_{-0.09}$ \\
$\mu_{L,\alpha*}$ [mas/yr] & $11.89^{+4.05}_{-3.90}$ & $12.05^{+4.46}_{-3.71}$ \\
$\mu_{L,\delta}$ [km/s] & $-7.76^{+3.77}_{-3.09}$ & $-5.24^{+3.28}_{-3.13}$ \\
$\theta_E$ [mas] & $2.60^{+0.21}_{-0.24}$ & $2.47^{+0.28}_{-0.24}$ \\
$\varpi$ [mas] & $0.28^{+0.05}_{-0.05}$ & $0.37^{+0.07}_{-0.06}$ \\
$\mu_{\alpha*}$ [mas/yr] & $-6.42^{+0.08}_{-0.06}$ & $-6.22^{+0.10}_{-0.08}$ \\
$\mu_{\delta}$ [mas/yr] & $1.44^{+0.08}_{-0.08}$ & $1.48^{+0.06}_{-0.07}$ \\
\end{tabular}
    \caption{Values inferred from the astrometric models corresponding to the two solutions found for the photometric model of GaiaDR3-ULENS-001 and astrometric parameters of the lensed source after including the microlensing effect.}
    \label{tab:models}
\end{table}

\begin{figure}
    \centering
         \centering
         \includegraphics[width=9cm]{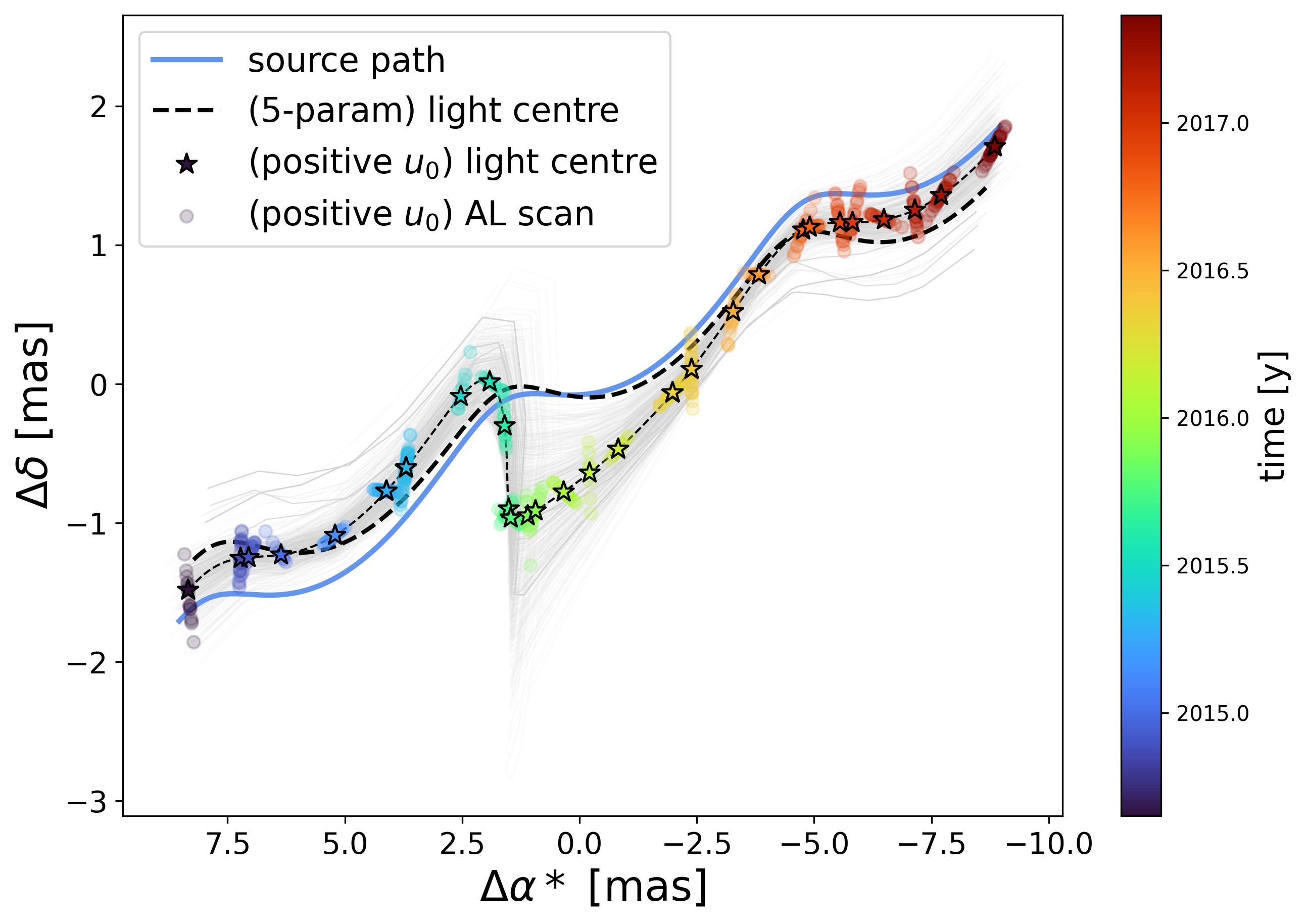}
    \caption{Simulated astrometric track for the GaiaDR3-ULENS-001 event. The solid blue line marks the source path without the microlensing effect. The thin, dashed black line marks the light centroid path. The coloured dots mark the light centroid positions during the measurement epochs per CCD. The thick, dashed black line shows the path corresponding to the 5-parameter fit obtained $Gaia$. The light gray lines represent randomly selected MCMC samples. The simulated track for the negative $u_0$ solution is very close and was omitted for clarity.}
    \label{fig:astrometry-track}
\end{figure}

\section{Discussion}
We have used the anomalous astrometric $Gaia$ DR3 measurement for GaiaDR3-ULENS-001 microlensing event in order to derive the size of its angular Einstein Radius. If real, it would be the very first astrometric microlensing measurement using only $Gaia$ observations. It relies on the assumption that the astrometric 5-parameter $Gaia$ model is a bad fit because of the microlensing effect distorting the source trajectory. At this stage, of course, we can not rule out the anomaly is just some kind of instrumental effect or even some different astrophysical effect than microlensing. On the other hand, observing an astrometric anomaly in a bright, isolated star which underwent a photometric microlensing event is very likely to be due to microlensing. 


The nature of the lens\footnote{Following a common convention the lensing body in this microlensing event should be denoted as GaiaDR3-ULENS-001L.} can be inferred using a combination of the inferred mass, its distance and the amount of light it emits, what can be estimated using blending parameter from the microlensing photometric model. 
Although the blending median values in both solutions concentrated around 1.0 (i.e. 100\% of lensed light comes from the source and the lens is dark), the distribution tails could  correspond to some amount of light from the blend. In principle, the blend in dense stellar regions could be also a third light source, not associated with the event, however here we explore the scenario that the blend light could be coming from the lens. We investigate if that light is enough for a most common type of lens in the Galaxy, namely a Main Sequence star. 

To estimate the probability of this outcome, we used a simulation that uses the expected luminosity of a Main Sequence star at the inferred distance to estimate the probability of the considered lens being a Main Sequence star assuming the whole blended light is the lens' light, following \cite{Wyrz16} and \cite{Mroz_Wyrzykowski_2021}. Having the astrometric model in addition to the photometric model, the relative motion value $\mu_{rel}$ can be derived. The positive $u_0$ solution yields a relative lens-source proper motion of $20.62^{+3.47}_{-2.88}$ mas/yr and  $20.04^{+4.56}_{-3.32}$ mas/yr for the negative $u_0$ solution.

Since the exact value of the extinction to the lens is not known, all the results for the extinction between 0 mag and 2.35 mag are considered \citep{EDR3}.
If there is no third light present, the chance of the lens being a dark lens depends greatly on the extinction. Extinction Galaxy model STILISM indicates the reddening for 0.90kpc (0.69kpc) is 0.17$\pm$0.07 mag (0.16$\pm$0.05 mag). If real, the probability for the lens not being a Main Sequence star is over 89.58\% (75.80\%) for positive (negative) $u_0$ solution \citep{STILISM}. 


The $1\sigma$ ranges of the inferred lens masses for both solutions fit well under the Chandrasekhar limit, therefore a massive white dwarf (WD) is a most likely possibility in case of a dark lens.

\begin{figure}
    \centering
    \includegraphics[width=9cm]{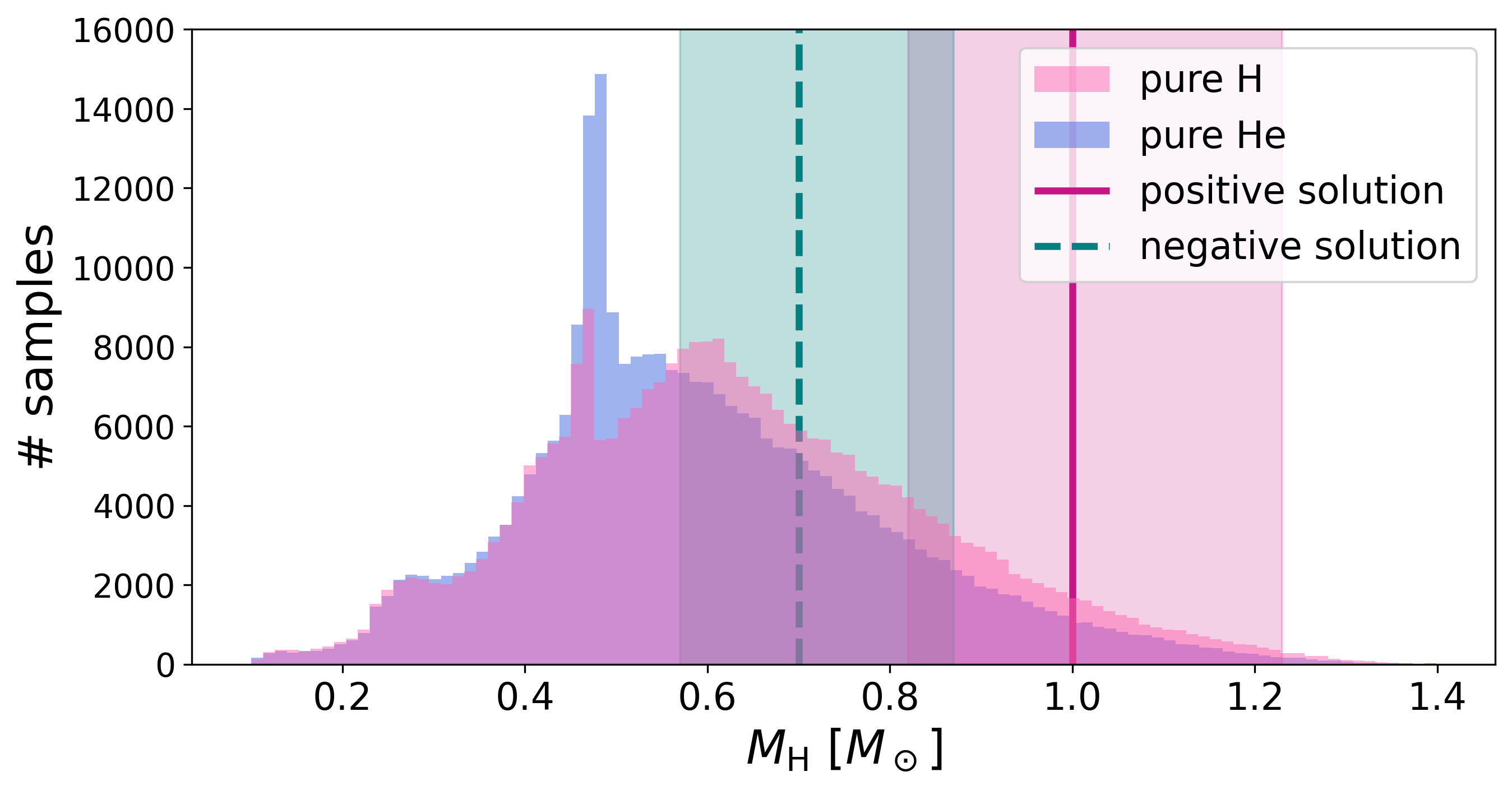}
    \caption{EDR3 white dwarfs' masses distributions for mass distributions resulting for two models assuming pure hydrogen composition and pure helium composition.}
    \label{fig:wd_gentile_mass_hist}
\end{figure}

The resulting mass was compared with the masses of white dwarfs found using $Gaia$ EDR3 by \cite{gentile2021catalogue}. Figure \ref{fig:wd_gentile_mass_hist} shows the distributions of masses of WDs with the assumptions of pure hydrogen and pure helium atmospheres. The negative $u_0$ solution for lens in GaiaDR3-ULENS-001 event would match an average-mass WD, while the positive $u_0$ solution -- a high-mass WD.


The high end of the mass distribution coincides with a few cases of small neutron star masses known \citep{lattimer_masses, ozel2016masses, rawls2011refined}. However, some theoretical models suggest the minimal mass to be around 1.17 $M_\odot$ \citep{suwa2018minimum}. Therefore, it is less likely, although not impossible, that  GaiaDR3-ULENS-001L is a low-mass neutron star.

The final conclusions on the nature of the lens should be possible in couple of years when the lens and the source separate on the sky. Given the relative proximity of the lens, it should be possible to confirm or reject Main Sequence or white dwarf scenarios. 

\section{Conclusions}
The measurement of the astrometric effect of microlensing together with the photometric observations is hoped to let unambiguously infer the parameters of lens mass, distance, and proper motions. 
The \textit{Gaia} mission delivers astrometric measurements with miliarcsecond precision, which is enough to observe astrometric microlensing caused by stellar-mass lenses. 

\textit{Gaia} Data Release 3 does not contain astrometric time-series, but just the 5-parameter astrometric solution. Here we modelled the event GaiaDR3-ULENS-001 using its photometric $Gaia$ lightcurve and values from the 5-parameter astrometric solution under the assumption that the large RUWE value is the result of light centroid deviations caused by microlensing. We found two photometric models and two corresponding astrometric models, using the photometric values as an input.

The inferred lens mass is $1.00^{+0.23}_{-0.18}$ $M_\odot$ and $0.70^{+0.17}_{-0.13}$ $M_\odot$ for positive and negative solution, respectively. Since the blending value is close to 1.0, meaning the lens is probably a low-luminosity or dark object, a white dwarf is the most likely explanation of the lens' nature. 
In principle, given the sparsity of the $Gaia$ light curve, we can not fully rule out the lens was a tight binary system and in such case the measured mass was the total mass of the system. 
Further detailed high-angular resolution imaging in years after the event are necessary to fully confirm the nature of the lens.

The methodology used here proved to be able to put much more specific constraints on the lens mass than based on the photometry alone. Future \textit{Gaia} data releases are also going to contain astrometric time-series, which will provide the mass measurements even further for hundreds of events. The large-scale hunt for isolated black holes may finally begin.


\section*{Acknowledgments}
We thank Drs. Ulrich Bastian, Berry Holl, Pawe{\l} Zieli{\'n}ski, Mariusz Gromadzki, Radek Poleski, Przemek Mr{\'o}z, Andrzej Udalski and the OGLE team. This work has made use of data from the European Space Agency (ESA) mission
{\it Gaia} (\url{https://www.cosmos.esa.int/gaia}), processed by the {\it Gaia}
Data Processing and Analysis Consortium (DPAC,
\url{https://www.cosmos.esa.int/web/gaia/dpac/consortium}). 
This work was supported from Polish NCN grants:  Harmonia No. 2018/30/M/ST9/00311 and Daina No. 2017/27/L/ST9/03221. We acknowledge the European Commission's H2020 grant ORP No 101004719.

\bibliography{bibs}
\bibliographystyle{aa}

\end{document}